# Characterization of Surface Deformation Behavior, Mechanical and Physical Properties of Modified-clay Bricks

David N Githinji* Charles K Nzila, John T Githaiga, David R Tuigong, Albert O Osiemo and Peter O Ayoro

Department of Manufacturing, Industrial & Textile Engineering, Moi University, P.O Box 3900-30100, Eldoret, Kenya

**Abstract**

The demand for building material is ever increasing owing to population growth. Compacted clay bricks are an important integral building material especially for low cost durable and affordable housing segment. This is a valued building material since its properties can be modified to suit various loading conditions. In this paper, the mechanical and physical properties of clay bricks modified with varying proportions of sawdust and polystyrene are determined. Increment of non-clay material proportion in the modified-clay bricks increases their porosity and water absorbency while their bulk densities, compressive and flexural strengths decreases. The use is made of Particle Image Velocimetry (PIV) method to assess the surface deformation behavior of the modified-clay bricks under uniaxial compressive loading. The distribution of surface deformation as assessed through PIV method is relatively uniform in pure-clay bricks while modified-clay bricks indicates a non-uniform deformation localized near the loading point at low strains. The strain distribution progressively spread out in the modified-clay brick as the failure point is approached.

**Introduction**

Clay is a natural material composed mainly of crystalline minerals which are essentially hydrous aluminium silicates with small quantities of iron or magnesium replacing aluminium partly or wholly in some minerals (Murthy 2002). Clay may also contain water-soluble minerals and organic material as discrete particles or as organic molecules adsorbed on clay minerals. Moist clay material exhibits plasticity and upon curing a rigid structure is formed. The clay mineral in the structure decomposes into free alumina, free silica and water vapour on heating. On further heating, a liquid glass is formed from some of these free oxides, alkalis and other fluxes in the clay material, which crystallizes into mullite at temperatures above 1200°C. The constituents of clay bricks therefore include mullite crystals, silica and a glassy matrix with minor constituents of unaltered quartz, free alumina, calcium and other silicates (Engineers 2007). As natural material, clay has varying chemical composition depending on changes in the environment where clay deposits are found. Consequently, the mechanical and physical properties of clay bricks vary on a regional level. Characterization of clay bricks is therefore relevant to understand their physical and/or mechanical properties so as to evaluate their safety in load bearing structural applications.

Most studies on the characterisation of modified-clay brick majors on their physical and mechanical properties. Production of lightweight bricks from clay and polystyrene has been reported by (Veiseh 2003). The study showed an increase in water absorption and a reduction in compressive strength and density as the fraction of polystyrene was increased. Similar results were reported for clay brick modified with varying proportion of sawdust (Chemani 2012). Information about characterisation of clay brick's deformation behavior using digital image analysis is rather scarce. Digital image correlation has been applied to study fracture toughness of clay brick (Lorenzo et al. 2014) and strain measurements of large masonry walls (Salmanpour and Mojsilovic 2013). This paper aims at applying digital image correlation method to assess the surface deformation behavior of modified-clay bricks. The Particle Image Velocimetry (PIV) plugin installed in the ImageJ software (Abramoff et al. 2004) is used. PIV is a displacement measuring technique of small particles embedded in a region of a fluid. The theory, working principle and applications of PIV is expounded in several studies (Westerweel 1993, 1997, 2000; Willert 1996; Willert and Gharib 1991). The method works by recording at different times, the light reflection from tracer particles in a flowing medium, thus allowing evaluation of their displacement. A typical PIV setup consists of a digital camera, light source and image analysis software. For accurate PIV assessment, the camera setting and lighting are maintained constant during the measurement. The particle movement from one point to another is quantified in terms of vector magnitude. The PIV method has been applied previously (Githinji et al. 2015) in strain assessment in other materials. The findings of the current work will be of great importance in structural integrity assessment of loaded modified-clay brick structures.

## Materials and Methods

### Materials

The clay soil used in this study was collected from the same ecological zone. The soil was sun-dried for 5 hours and passed through a sieve having 3.25 mm apertures. The sun-dried sawdust and polystyrene materials were manually shredded and sieved through 3.25 mm apertures.

### Methods

#### Fabrication of Clay Bricks

Pure-clay bricks were fabricated by adding water slowly to the clay soil while continuously mixing until a plastic paste was formed. The paste was poured into a steel mould measuring 190 mm x 90 mm x 65 mm followed by compacting in a hydraulic compressing machine for 5 minutes at 2.5 MPa. The moulded bricks were dried at 100°C for 6 hours under atmospheric humidity before firing at 1000°C for 18 hours in a furnace. At the end of firing, the furnace was kept closed and slow cooling allowed for about 12 hours. Modified-clay bricks were prepared by mixing water, clay, sawdust and polystyrene at different ratios as summarized in Table 1. For each category of brick, 5 samples were fabricated under similar conditions and used in the current study. Only average values are reported in this work.

Table 1: Fabricated brick components based on volume ratio

| Sample | Pure Clay | Polystyrene | Sawdust |
|--------|-----------|-------------|---------|
| A      | 100%      | 0           | 0%      |
| B      | 25%       | 25%         | 0%      |
| C      | 50%       | 50%         | 0%      |
| D      | 75%       | 75%         | 0%      |
| E      | 25%       | 0%          | 25%     |
| F      | 50%       | 0%          | 50%     |
| G      | 75%       | 0%          | 75%     |

#### Characterisation of Physical Properties of Fabricated Clay Bricks

The average density of the fabricated bricks was computed from their average mass and average volume. The measurement of dimensions was in accordance with Kenya Standard KS EAS 54:1999 (KeBS 1999). The total volume of porosity was computed from the weight gain of oven-dried clay bricks saturated with water (Duggal 2009). The drying was at 100°C for 24 hours while water immersion test was performed for a similar duration. The void volume fraction was given by the ratio of total-void volume to saturated-brick volume. The volume difference between saturated-brick and oven-dry brick, gave the total-void volume. The determination of water absorption was in accordance with the test method specified in Kenya Standard KS EAS 54: 1999.

#### Characterisation of Mechanical Properties of Fabricated Clay Bricks

The compressive strength of the bricks was determined in accordance with Kenya Standard KS EAS 54: 1999. The dry compressive tests were conducted on a Uniaxial Testing machine. The tests were performed in displacement controlled mode at a constant machine crosshead rate of 2.5 mm/min on specimen measuring 30 mm x 30 mm x 50 mm extracted in the moulding direction from the fabricated bricks. The specimen contact surfaces with machine platen were grounded to allow proper alignment and uniform loading during the test. The specimen's height/width ratio was more than 1.5 which was adequate for minimizing the boundary effect during the test. The compressive strength was calculated by dividing the peak compressive load by the original cross-sectional area of the brick specimen. The flexural rigidity test was performed on a Uniaxial Testing machine following a 3-point bending test on the fabricated brick samples. The flexural rigidity ($\delta$) was calculated from the maximum load applied (F) before failure and the Length (L), width (W), and height (H) of the fabricated brick as: $\delta = 1.5(FL/BH^2)$.

**Characterisation of Surface Deformation Behavior of Fabricated Clay Bricks**

A digital camera (Nikon D3100 14.2 MP with 18-55 mm VR Lens) was used to capture sequential images of the brick specimen during the compressive tests. The assessment was carried out on pure-clay brick (sample A), polystyrene-clay brick (sample C) and sawdust-clay brick (sample F). For every 0.5 mm crosshead displacement, an image of the brick surface was captured and this was repeated until the specimen fractured. ImageJ software installed with a PIV plugin was used to qualitatively analyze the uniformity of surface deformation, using vector magnitude maps. This was achieved by correlating an image captured at zero strain with an image captured at a higher strain level. The strain built-up, after each 0.5 mm compression interval, was quantified by averaging the vector magnitudes of all elements displacement in the camera field of view.

**Results and Discussion**

**Characterisation of Fabricated Clay Bricks Properties**

The experimental results for the different characteristics of the fabricated brick samples tested that include average values of void fractions, water absorptions, bulk densities, compressive strength, flexural strength and loss of ignition are given in Table 2.

Table 2: Characterized properties of modified-clay bricks (S- and P- denote sawdust and polystyrene, respectively)

| Sample | Clay Fraction | Void fraction | % water absorption | Bulk density (kg/m$^3$) | Compressive Strength (MPa) | Flexural Strength (MPa) |
|---|---|---|---|---|---|---|
| A | 100% | 0.12 | 14 | 1626 | 2.31 | 0.21 |
| P-B | 25% | 0.43 | 40 | 463 | 0.54 | 0.05 |
| P-C | 50% | 0.24 | 26 | 729 | 0.65 | 0.08 |
| P-D | 75% | 0.17 | 18 | 1468 | 1.58 | 0.11 |
| S-E | 25% | 0.29 | 75 | 541 | 0.69 | 0.07 |
| S-F | 50% | 0.21 | 14 | 771 | 0.82 | 0.12 |
| S-G | 75% | 0.16 | 9 | 1519 | 1.73 | 0.15 |

**Properties of Pure-clay Brick**

The determined void volume fraction value of 0.12 was found to be within the nominal range (0.01-0.5) for the clay bricks as reported by other researchers (Engineers 2007; Phonphuak 2013). A relatively high water absorption value of 14% obtained may be attributed to the void volume fraction which provided space for water during the immersion tests. Moreover, the value is comparable to the reported values in literature (Cobb 2008) for similar clay bricks. The pure-clay bricks bulk density was good since the determined quantity (1636 kg/m$^3$) was within the typical range (1600-1800 kg/m$^3$) for clay bricks previously reported (Duggal 2009), which mainly depends on clay composition and void volume fraction (Phonphuak 2013). For the determined brick compressive strength of 2.3 MPa, a closely correspondence to that specified in the KS EAS 54:1999 Standard for internal clay bricks was established. The computed flexural strength was only about 7-10% of the compressive strength which may be ascribed to the relatively low tensile strength of brittle material. Comparable flexural strength values have been reported previously (Cobb 2008) for similar clay bricks. It is also possible that the clay bricks tested had lines of weakness owing to agglomeration of voids in the structure, which may explain, why crack initiation and development originated along specific areas of the brick (see Fig. 2). Since the physical and mechanical properties reported in this paper for the pure-clay brick correspond to those in published literature, then it provide a good reference for studying the effects of clay composition variations on these properties.

The installed software in the Universal Testing Machine, allows for the results to be presented in various formats of deformation. The nature of the failure and related stress-strain behavior is continuously computed during the compression loading cycle. Fig. 1 shows the engineering stress-strain curve for the pure clay brick sample loaded in compression to failure. From the results, it is clear that the failure occurred in a brittle manner with negligible plastic strain. This behavior could be attributed to the strong bonding between clay particles (e.g. various oxides of Si, Ca, Ti, Al) after high temperature curing resulting in rigid structures. The accumulated compressive strain on loading can be attributed to relative displacement between different parts of cracked brick.

Since the brick is brittle in nature, this compressive strain is accommodated by the brick's lateral spread which leads to crack widening with increasing strain. The stress-strain curve's none-linearity can be attributed to occasional slight reduction in load bearing capacity of the brick owing to cracking with the subsequent increase in load bearing arising from redistribution of stress in the structure upon further compression.

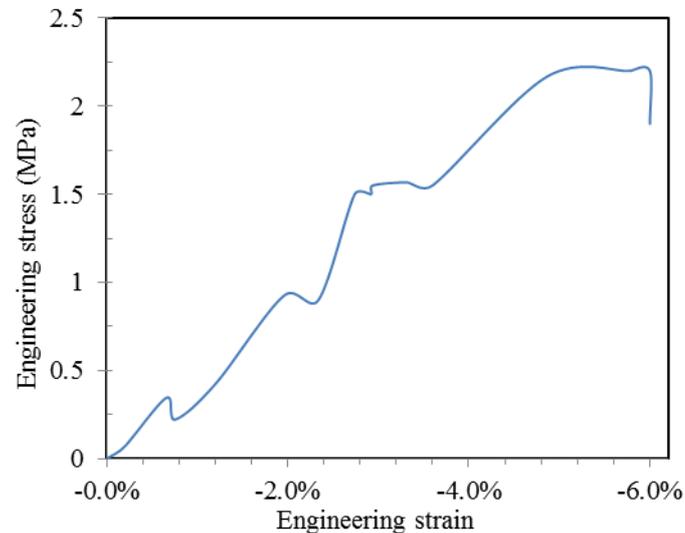

Fig. 1. Compressive stress-strain curve for pure clay brick loaded to failure.

**Properties of Polystyrene Modified-clay Brick**

The void volume fraction of the modified-clay brick increased with the proportion of polystyrene used. This may be attributed to decomposition of polystyrene at the curing temperature into gaseous by-products (Wunsch 2000), which may have escaped from the brick structure creating voids (see Fig. 2). Correspondingly, absorption of water increased with increase in polystyrene proportion owing to large number of voids. The increased void fraction may also have reduced bulk density as the proportion of polystyrene was increased. A progressive reduction in the compressive strength occurred as the ratio of polystyrene was increased. This is in agreement with previous studies (Veiseh 2003) and may be ascribed to the presence of voids in the modified-clay brick structure. This may have thus reduced the effective cohesion between the clay particles due to increased inter-particle spaces, thus lowering the compressive and flexural strengths of the brick. It is therefore possible that the amount of polystyrene used determines the size of voids produced in the structure. With references to pure-clay brick, the slight reduction in bulk density (by about 10%) and considerable reduction of compressive strength (by about 30%) observed in the modified-clay brick with 75% clay and 25% polystyrene, may be ascribed to the presence of many minutes voids.

**Properties of Sawdust Modified-clay Brick**

The void volume fraction of the modified-clay brick increased with the proportion of sawdust used. This may be attributed to decomposition of wood at the curing temperatures into gaseous and solid by-products (ashes) (Chaula 2014), where escaping gases creates voids in the structure. Correspondingly, absorption of water increased with the sawdust proportion owing to increased voids fraction. The relatively high water absorption for bricks containing 75% sawdust may be as result of water absorption and retention by the ashes and remnant sawdust in the brick. The water absorption of sawdust has been reported to be relatively high owing to high porosity (Sciences 1999). A progressive reduction in the compressive strength occurred as the ratio of sawdust was increased. This agrees well with other studies (Chemani 2012) and may be attributed to the presence of voids as explained previously. The relevantly low reduction in bulk density as compared to pure-clay brick may be attributed to the presence of ashes in the brick following the decomposition of sawdust.

**Comparison between Modified-clay Bricks Properties**

Modified clay bricks with similar ratios of sawdust and polystyrene had distinct properties, with addition of sawdust giving superior mechanical properties. This may be attributed to relatively high thermal composition temperature of wood (about 500°C) (Gašparovič 2009) compared to that of polystyrene (about 100°C) leading to high void fractions and thus relatively low mechanical properties in the polystyrene bricks than in the sawdust bricks. The presence of ashes and non-decomposed wood components in the sawdust brick may explain the high water absorption observed as these components have capacity to absorb and retain water.

**Surface Deformation Behavior of Modified-clay Bricks**

The results of the surface deformation behavior of the modified-clay bricks through image measurements and analysis are presented. Fig. 2 shows vector magnitude maps generated from sequential macro-images taken at an engineering compressive strain of 2% and at failure. The vector magnitude maps were obtained by correlating the image captured at 0% strain with those captured at higher engineering strains. The degree of displacement of surface elements as tracked progressively from their original positions is represented by a rainbow coloring scale. The blue colour indicates the least displacement while the red colour designates the maximum displacement of the surface elements.

From Fig. 2, it is evident that the displacement at the surface progressed along certain regions of the pure-clay brick, where cracks initiated and developed as loading was increased. This may represent line of weakness in the manufactured clay brick where fusion of clay particles is inadequate. The accumulated compressive strain (see Fig. 1) could therefore be attributed to this displacement arising from cracking rather than the plastic deformation of the brick which had failed in a brittle manner.

The maximum surface displacement in the modified-clay bricks was localized around the loading regions while displacement in centre region of bricks was nominal. This indicates limited load transfer in the modified-clay bricks which may be attributed to relatively high void fraction (~ twice that of pure-clay brick). During loading, the voids collapse progressively starting from the outer surface toward the brick's centre. This trend is qualitatively demonstrated by the vector magnitude maps (see Fig. 2) for the modified-clay bricks, where maximum displacement at low strain is indicated at bricks' edges but as the strain increases, the displacement spreads toward the centre of the brick. Interestingly, limited macro-cracks developed in the modified-clay brick before failure. This underpins the importance of clay brick modification for structural application requiring some resilience and less dense materials.

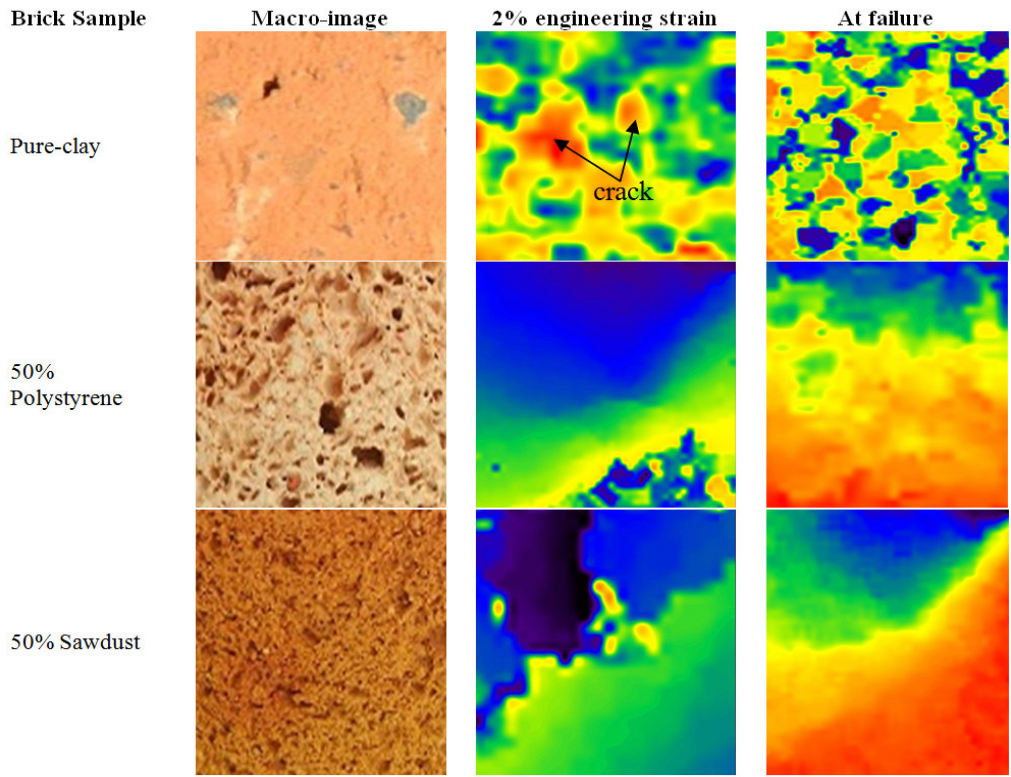

Fig. 2. Macrographs and their corresponding vector magnitude maps for: pure-clay brick, 50% polystyrene-clay brick, and 50% sawdust-clay brick. Rainbow colour coding from blue to red represent minimum to maximum strains in the structure.

**Conclusion**

In the current study, clay bricks were fabricated using different proportions of sawdust and polystyrene material and their physical and mechanical properties characterized. The bricks' surface deformation behavior was evaluated using Particle Image Velocimetry (PIV) method and the following conclusions drawn from the study:

- The determined physical properties (void fraction, bulk density, and water absorption) and mechanical properties (compressive and flexural strength) of the pure-clay bricks fabricated in this work, agrees well with the results reported in literature and hence the samples are a fair representative of this type of building material.
- The stress-strain curve of pure-clay bricks exhibit brittle failure thought to be as result of strong bonding between clay particles on high temperature curing.
- The modification of clay bricks with increasingly high proportion of polystyrene and sawdust progressively increases bricks' void fraction and water absorption while lowering their compressive and flexural strengths. The reduction of effective cohesion between the clay particles due to increased inter-particle spaces is thought to have negative impact on the bricks' strengths.
- The strain distribution in the modified-clay bricks as assessed through PIV method is non-uniform and localized near the loading point at low strains. The strain distribution progressively spread-out in the brick structure as the failure is approached. This is thought to be as result of increased voids which collapse gradually thus allowing load advancement through the structure.

**Acknowledgement**

The authors would like to thank VLIR-OUS MU for partially funding this research.